# Testing the performance of Multi-class IDS public dataset using Supervised Machine Learning Algorithms

Vusumuzi Malele[1] and Topside E Mathonsi[2]

[1,2]Department of Information Technology, Faculty of Information and Communication Technology, Tshwane University of Technology, Pretoria, South Africa
`Vusimalele@gmail.com` and
`MathonsiTE@tut.ac.za.ac.za`

**Abstract.** Machine learning, statistical-based, and knowledge-based methods are often used to implement an Anomaly-based Intrusion Detection System which is software that helps in detecting malicious and undesired activities in the network primarily through the Internet. Machine learning comprises Supervised, Semi-Supervised, and Unsupervised Learning algorithms. Supervised machine learning uses a trained label dataset. This paper uses four supervised learning algorithms Random Forest, XGBoost, K-Nearest Neighbours, and Artificial Neural Network to test the performance of the public dataset. Based on the prediction accuracy rate, the results show that Random Forest performs better on multi-class Intrusion Detection System, followed by XGBoost, K-Nearest Neighbours respective, provided prediction accuracy is taken into perspective. Otherwise, K-Nearest Neighbours was the best performer considering the time of training as the metric. It concludes that Random Forest is the best-supervised machine learning for Intrusion Detection System.

## 1    Introduction

The trend in the Information and Communication Technology (ICT) sector leads to data and information being stored and accessed from anywhere. For example, a mobile service technician can generate data, save it, and access it using Internet-of-Things (IoT) and cloud computing platforms. He/she could later use that data or use that of the other colleagues. The IoT is the network of physical objects (i.e. sensors, actuators, controllers, computers, etc) accessed through the Internet. While cloud computing is platform that allows for the storing and accessing of data over the Internet, instead of using the local computer hard drive.

   The IoT and cloud computing trends expose organisations to network and data security vulnerabilities and risks. Since, data, information and knowledge form major parts of the organisations' assets, it needs to be protected. Any leakages or unauthorized sharing and access of critical data should be picked up immediately to avoid putting the organisation into serious business challenges.Vulnerabilities and risks in IoT and cloud computing affect the organisations' confidentiality, availability and integrity of offering services.

    Avoiding data leakage in these days of IoT and cloud computing is important. In this case, Intrusion detection systems (IDS) could assist. The IDS is a software that helps in detecting malicious and undesired activities in the network primarily through the Internet. In this regard, IDS is a necessary solution for all organizations. The IDS detects various types of attacks and it is categorized into two groups: (i) Signature-based Intrusion Detection System (SIDS), and (ii) Anomaly-based Intrusion Detection System (AIDS) [1]. AIDS overcomes the shortcomings of SIDS and it could be implemented using any of the following three methods [1]: (a) machine learning, (b)

statistical-based and knowledge-based methods.

Machine learning is a branch of artificial intelligence (AI). It is the process of extracting decision-making information from a large set of data. To recognise or predict behaviour and/or determine data patterns, machine learning uses the set of rules, methods, and/or functions [2]. Machine learning comprise three broad learning algorithms, Supervised Learning, Semi-Supervised Learning, and Unsupervised Learning. Supervised machine learning uses a trained label dataset, Unsupervised Learning does not use the labeled dataset and Semi-Supervised Learning uses small label dataset to guide huge unlabelled datasets. This paper uses four supervised learning algorithms to test the performance of the public dataset. The algorithms are Random Forest, XGBoost, K-Nearest Neighbors (k=5), and Artificial Neural Network (ANN).

The remaining part of this paper is organized as follows: the next section will briefly discuss the literature review, followed by the methodology, then results and findings will be presented, and the last section will provide the conclusion and future work.

## 2    Literature Review

This section discusses the literature that relates to this work. It begins by briefly looking at machine learning algorithms, then network attacks that could be resolved through machine learning and conclude by summarising the related work.

*A.   Supervised Machine Learning*

The are several algorithms used to implement Supervised Machine Learning. The focus of this paper is on Random Forest (RF), eXtreme Gradient Boosting (XGB) also known as XGBoost, K-Nearest Neighbors (k=5) (KNN), and Artificial Neural Network (ANN).

- Random Forest (RF): comprise building many decision trees that are uncorrelated [3]. Many decision trees create an efficient method for estimating unlabelled data with very high performance in addition to regression problems.
- XGBoost (XGB): uses a gradient boosting algorithm, known as boosted tree algorithm. It has a very high speed in addition to a significant performance [4]. Thus, it is dominating any other algorithm for structured data.
- K-Nearest Neighbours (k=5) (KNN):  used for large datasets and low dimensions [5]. The value k= 5 means that any data point is classified based on its nearest five neighbours classification, based on neighbours voting. For example, if four out of five neighbours belong to class A, then the decision will be to make the last one to belong to class A.
- Artificial  Neural Network (ANN): comprise fault-tolerance and abrupt response and it is an artificial intelligence technique used to solve rigorous problems, which humans are not able to.

*B.   Machine Learning Attacks*

As an AI area, machine learning is used to focus on studying network traffic, learning the types of incoming traffic, and detecting any anomalous behaviour in the network. To build an intelligent security system, it is essential to increase the alertness towards network malicious behaviours or attacks. Network attacks are broadly classified into three different categories [5, 6]:

- Denial-of-Service (DoS) Attacks: these attacks seek to slow down the traffic in a network, and sometimes, this could result in shutting down the whole network. Some kinds of denial of service attacks are flooding DoS, distributed DoS, and flaw exploitation DoS.
- Penetration Attacks: these attacks aim to gain access to the whole system without authorization from the network administration; consequently,  the attacker can have access to the system files and change states of the network. The common penetration attacks are: User-to-Root (U2R), Remote-to-Root (R2R), Remote-to-User (R2U), and Remote Disk Read (RDR) attacks.
- Reconnaissance or Scanning Attacks whose aim is to scan networks, ports, vulnerabilities, etc., to gain information about the whole network topology, identify the computer users, firewall types into the system, and the operating systems. It subdivided into two [6]: Logical Reconnaissance (i.e. anything that is done in the digital spectrum and in most cases, network admin does not have control [6]) and Physical Reconnaissance (i.e. crosses the lines of what in this case, a network admin has control of issues; however, there are elements that will never be fully protected).

*C. Similar Work*

Machine learning could be viewed as the most effective technique for IDS. It should be used in protecting people and business integration that happens over the network/internet. In this regard, different work has been conducted looking at its performance. For example:

- the study by [7] looks at machine learning in the context of IoT. It showed that machine learning is efficient in detecting attacks; hence, making it relevant in IoT cybersecurity-related research. Another research work by [8] looked at IDS in IoT infrastructures using supervised learning algorithms to identify cyber-attack). They have applied the algorithms and showed their performances using two public datasets;
- the work by [9] contributed a new intrusion detection framework based on the feature selection and ensemble learning techniques. They created an ensemble classifier that achieved 81.31% of accuracy;
- while [10] build a classifier and proposed a new joint optimization algorithm using swarm intelligence to optimize the deep belief network (DBN) network structure for optimization. Their classifier work achieved a 82.36% highest accuracy;
- in [11] an ANN IDS was used on a public dataset. Their results showed promising performance on detecting malicious attacks in the network on real-time basis. Their performance evaluation was based on specific network features using the four essential prediction accuracy measures: true positive, true negative, false positive, and false negative; and
- the work by [12] classified the network traffic as normal or anomaly, and used machine learning techniques, k-nearest neighbours, decision tree, and support vector machine to evaluate IDS. Their evaluation used four necessary measures: accuracy, precision, sensitivity, and F1-score.

This paper used the DoS, U2R, Probing, and Root-to-local (R2L) as network malicious behaviours or attacks and the RF, XGB, KNN, and ANN as supervised machine learning techniques to compute the classification models. This paper adopted accuracy, precision, and F1-score. Furthermore, it used the true positive, true negative, false positive, and false negative as the accuracy measure.

## 3 Research Approach

This section presents the methodology that was followed in conducting this study.

*A. Methodology*

Fig. 1 illustrates the methodology that was adopted to conduct this study. The public file dataset, Table 1, is included as an input, then taken through pre-processing. In pre-processing step the feature selection technique is used since the number of in attributes in the dataset is high. Pre-processing step allows all the categorical data which are in textual form be converted into numerical form using the feature selection techniques. Pre-processed data is divided into testing data and training data. The used features are summarised in Table 2.

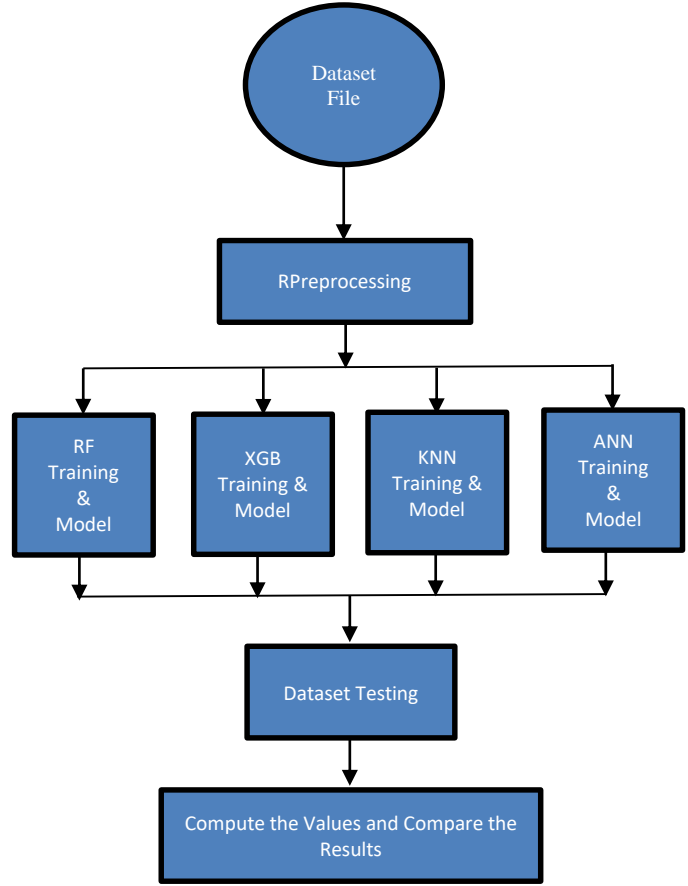

**Fig. 1.** The adopted methodology (*Author's adaptation*)

Subsequently, the output of the pre-processed step and the classification models/algorithm (RF, XGB, KNN, and ANN) were used to compute the classification models' results using some performance measures which are:

1) *the confusion matrix that consists of the following classification attacks:*

   - Denial-of-service (DoS): number of denial of services attack cases.
   - Probe: number of surveillance and probing cases.
   - Root-to-local (R2L): number of unauthorized access from the remote machine to local machine cases.
   - User-to-root (U2R): number of unauthorized access to local superuser privileges by a local unprivileged user cases.

2) *The four other performance measures are: precision, recall, F1-score, and accuracy are calculated based on the following equations:*

$$Precision = \frac{t_p}{t_p + f_p} \quad (1)$$

$$Recall = \frac{t_p}{t_p + f_n} \quad (2)$$

$$F1 - score = \frac{2 * Recall * Precision}{Recall + Precision} \quad (3)$$

$$Accuracy = \frac{t_p + t_n}{t_p + t_n + f_p + f_n} \quad (4)$$

Where $t_p$ = true-positive, $t_n$ = true negative, $f_p$ = false-positive, and $f_n$ = false-positive,

In summary, using the research approach in Fig. 1, the following algorithm was adopted:

3) Pre-process the data set.
4) The data set is divided as training data and testing data.
5) Build the classifier model on training data for(RF, XGB, KNN, and ANN)
6) Read the test data
7) Test the classifier models on training data
8) Compute and compare Normal, DoS, Probe, R2L, and U2R for all models.

B. *Public Dataset*

This paper focuses on a public dataset that was captured for IDS. The file contains 88,180 records that are labelled as normal or attack. Normal records are for non-malicious incoming network traffic and the attack records are for malicious incoming network traffic. The attack records are classified as DoS, Probe, R2L, and U2R. The public dataset is presented in Table 1 and its features are explained in Table 2. The 70% of the dataset is arranged into a training set of 70,459 samples and the remaining a testing set of 17,723 samples (30%).

C. *Hyper-parameters*

The hyper-parameters set for each algorithm are as follow:

- RF: 100 trees are built with depth of 13. The minimum sample is 1 and the size of the hyper-parameter search is 2.
- XGB: the algorithm uses 36 trees with maximum depth of 3.
- K-Nearest Neighbours: number of folds is k = 5. The Euclidean distance, p = 2, measures the true straight line distance between two points in Euclidean space.
- ANN: The number of layer is 10 Layers and rectified Linear Unit (relu) as activation function, and "Adam" solver. The learning rate is Alpha = 0.001.

**Table 1.** The Public Dataset

| A | B | C | D | E | F | G | H | I | J | L |
|---|---|---|---|---|---|---|---|---|---|---|
| 17 | 9 | 491 | 0 | 0 | 0 | 1 | 0 | 0 | 0 | N |
| 42 | 9 | 146 | 0 | 0 | 0 | 0.08 | 0.15 | 0 | 0 | N |
| 47 | 5 | 0 | 0 | 0 | 1 | 0.05 | 0.07 | 0 | 1 | D |
| 21 | 9 | 232 | 8153 | 0 | 0.2 | 1 | 0 | 0.04 | 0.01 | N |
| 21 | 9 | 199 | 420 | 0 | 0 | 1 | 0 | 0 | 0 | N |
| 47 | 1 | 0 | 0 | 0 | 0 | 0.16 | 0.06 | 0 | 0 | D |
| 47 | 5 | 0 | 0 | 0 | 1 | 0.05 | 0.06 | 0 | 1 | D |
| 47 | 5 | 0 | 0 | 0 | 1 | 0.14 | 0.06 | 0 | 1 | D |
| 49 | 5 | 0 | 0 | 0 | 1 | 0.09 | 0.05 | 0 | 1 | D |
| 47 | 5 | 0 | 0 | 0 | 1 | 0.06 | 0.06 | 0 | 1 | D |
| 47 | 1 | 0 | 0 | 0 | 0 | 0.06 | 0.06 | 0 | 0 | D |
| 47 | 5 | 0 | 0 | 0 | 1 | 0.02 | 0.06 | 0 | 1 | D |
| 21 | 9 | 287 | 2251 | 0 | 0 | 1 | 0 | 0.03 | 0 | N |

| A | B | C | D | E | F | G | H | I | J | L |
|---|---|---|---|---|---|---|---|---|---|---|
| 17 | 9 | 334 | 0 | 0 | 0 | 1 | 0 | 0.2 | 0 | R |
| 34 | 5 | 0 | 0 | 0 | 1 | 0 | 0.06 | 0 | 1 | D |
| 36 | 5 | 0 | 0 | 0 | 1 | 0.17 | 0.05 | 0 | 1 | D |
| 21 | 9 | 300 | 13788 | 0 | 0.11 | 1 | 0 | 0.02 | 0 | N |
| 11 | 9 | 18 | 0 | 0 | 0 | 1 | 0 | 1 | 0 | P |
| 21 | 9 | 233 | 616 | 0 | 0 | 1 | 0 | 0.03 | 0 | N |
| 21 | 9 | 343 | 1178 | 0 | 0 | 1 | 0 | 0.04 | 0 | N |
| 33 | 5 | 0 | 0 | 0 | 1 | 0.1 | 0.05 | 0 | 1 | D |

**Table 2.** Description of Public Dataset as presented in Table 1

| Features | Type | Name | Description |
|---|---|---|---|
| A | Integer | Service | the service of the destination network used. |
| B | Integer | Flag | indicates the status of the connection whether it is normal or not. |
| C | Integer | Src bytes | the number of bytes of the data transferred from source to destination in a single connection. |
| D | Integer | Dst bytes | the number in bytes of the data bytes transferred from destination to source in a single connection. |
| E | Integer | Root shell | a value that represents whether a root shell is obtained or not. It has a value of 1 if it is obtained and 0 otherwise. |
| F | Integer | Srv error rate: | the percentage of connections that have activated the flag s0, s1, s2, or s3, among the connections aggregated in srv count. |
| G | Integer | Same srv rate: | the percentage of connections that were to the same service, among the connections aggregated in the count. |
| H | Integer | Diff srv rate: | a percentage that refers to the connections that have services different than the ones aggregated in Count. The percentage of connections that were too different services, among the connections aggregated in count. |
| I | Decimal | Dst host diff srv rate: | a percentage of connections that were to different services, among the connections aggregated in dst host count (the number of connections having the same IP address for the destination host. This value is not used as a feature but used to find other features). |
| J | Decimal | Dst host srv serror rate: | the percentage of connections that have activated the flag s0, s1, s2, or s3, among the connections aggregated in dst host srv count (number of connections having the same port number). |
| L | Text | Label | the target value to be predicted. It can have one of four values: N-Normal, D-DoS, R-R2L, U-U2R, or P-Probe. |

## 4 Results and Discussion

In This section presents and discusses the results in order to conclude which best algorithm that can predict an attack and be used with IDS. After applying the dataset on each of the algorithms: RF, XGB, KNN, and ANN the confusion matrices are produced and represented in Tables 3, 4, 5, and 6. The rows represent the instances in an actual class and the columns represent the instances in a predicted class.

**Table 3.** RF confusion matrix

|        | Normal | DoS  | Probe | R2L | U2R |
|--------|--------|------|-------|-----|-----|
| Normal | 9386   | 3    | 12    | 11  | 0   |
| DoS    | 1      | 6512 | 2     | 0   | 0   |
| Probe  | 5      | 4    | 1649  | 0   | 0   |
| R2L    | 5      | 0    | 0     | 129 | 0   |
| U2R    | 4      | 0    | 0     | 0   | 0   |

**Table 4.** XBG confusion matrix

|        | Normal | DoS  | Probe | R2L | U2R |
|--------|--------|------|-------|-----|-----|
| Normal | 9030   | 10   | 85    | 198 | 89  |
| DoS    | 2      | 6503 | 7     | 3   | 0   |
| Probe  | 18     | 6    | 1629  | 2   | 3   |
| R2L    | 1      | 0    | 1     | 132 | 0   |
| U2R    | 1      | 0    | 0     | 0   | 3   |

**Table 5.** KNN confusion matrix

|        | Normal | DoS  | Probe | R2L | U2R |
|--------|--------|------|-------|-----|-----|
| Normal | 9351   | 7    | 33    | 20  | 1   |
| DoS    | 9      | 6502 | 4     | 0   | 0   |
| Probe  | 46     | 11   | 1600  | 1   | 0   |
| R2L    | 17     | 0    | 1     | 116 | 0   |
| U2R    | 4      | 0    | 0     | 0   | 0   |

**Table 6.** ANN confusion matrix

|        | Normal | DoS  | Probe | R2L | U2R |
|--------|--------|------|-------|-----|-----|
| Normal | 9243   | 66   | 92    | 11  | 0   |
| DoS    | 245    | 6237 | 33    | 0   | 0   |
| Probe  | 213    | 38   | 1407  | 0   | 0   |
| R2L    | 102    | 0    | 22    | 10  | 0   |
| U2R    | 4      | 0    | 0     | 0   | 0   |

**Table 7.** ANN confusion matrix

|     | Precision (%) | Recall (%) | F1-score (%) | Accuracy (%) |
|-----|---------------|------------|--------------|--------------|
| RF  | 78.2          | 79.1       | 78.6         | 99.7         |
| XGB | 76.3          | 76.4       | 76.3         | 99.1         |
| KNN | 67.3          | 93.5       | 71.3         | 97.6         |
| ANN | 66.2          | 57.3       | 58.7         | 95.3         |

Table 7 presents the evaluation metrics that indicate the performance results for the different classification models used in this study. The RF model has the best performance accuracy of 99.7% and the higher F1-score value of 78.6%. The XGB model has the second-best performance accuracy and F1-score results computed to be 99.1% and 76.3%, respectively. It could be concluded that both RF and XGB are best algorithms to be used with IDS. The latter are compared with the findings of [1] who created an ensemble classifier that achieved 81.31% of accuracy. As well as that of [10] which reached the highest accuracy of 82.36%. Of note is that the KNN is also good for being used with IDS as it yields a 97.6% accuracy. In this paper, RF and XGB are best it was necessary to present the features (i.e. drivers of the results obtained due to their significant impacts on the output values) that have the most predictive power. The features scores are presented as shown in Fig. 2.

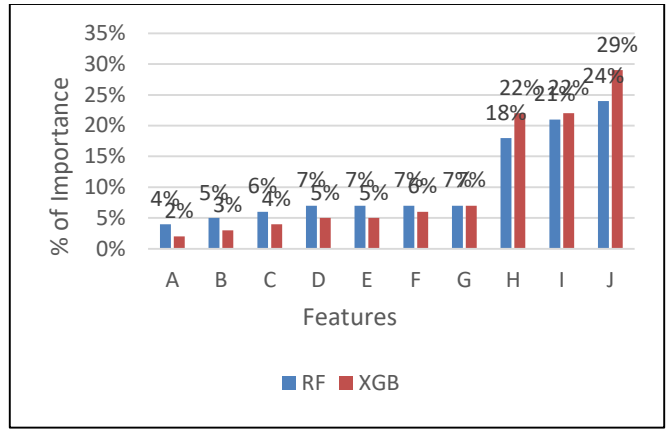

**Fig. 2.** The RF and XGB percentage of variable importance

**Table 8.** ANN confusion matrix

|  | **RF** | **XGB** | **KNN** | **ANN** |
|---|---|---|---|---|
| **Training Time in seconds** | 37 | 36 | 17 | 41 |

Training time is an important metric to consider when choosing which algorithm to use in IDS. Table 8 shows the training time for each algorithm. However, Clear the KNN with 17 seconds is the best followed by Both RF and XGB that an acceptable training time of around 37 seconds. Unfortunately, the accuracy of KNN is below that of RF and XGB, making it not be the best performing algorithm.

## 5   Conclusion and Future Work

This paper discusses the classification problem of IDS using supervised machine learning algorithms (RF, XGB, KNN, ANN). A publicly available dataset that contains common attacks (DoS, Probe, R2L, and U2R) was used. Based on the prediction accuracy rate, the results show that RF performs better on multi-class IDS, as well as XGB and KNN taking second and third place, respectively. However, if time of training could key metric that is considered, then KNN could have been preferred over RF and XGB.

In the future, this study will concentrate on evaluating the performance of the dataset using other supervised learning algorithms such as Logistic Regression (LR), and Support Vector Machine (SVM). Furthermore it will look at other performance measures such as Log Loss (an error metric that considers the predicted probabilities, which indicate the lower the log loss value is, the better the algorithm performance is).

**Acknowledgments**

The authors would like to thank the Tshwane University of Technology for financial support. The authors declare that there is no conflict of interest regarding the publication of this paper.